\documentclass[10pt,english,oneside,twocolumn,a4paper]{article}
\usepackage{graphicx}
\usepackage[utf8]{inputenc}
\usepackage{times}
\fontfamily{ptm}\selectfont
\usepackage{booktabs}
\usepackage{ragged2e}
\usepackage[singlelinecheck=false]{caption}
\usepackage[T1]{fontenc}
\usepackage[numbers]{natbib}
\usepackage{amsmath}
\usepackage{amssymb}

\usepackage{siunitx}
\usepackage{subcaption}
\usepackage{cleveref}
\crefname{equation}{}{}
\crefname{figure}{figure}{figures}

\usepackage{pgfplots}
\pgfplotsset{compat=1.15}
\usepgfplotslibrary{units}
\usepgfplotslibrary{fillbetween}
\usepgfplotslibrary{external}
\usepgfplotslibrary{colormaps}
\pgfplotsset{every axis/.append style={
		label style={font=\small},
		tick label style={font=\small, /pgf/number format/fixed},
		legend style={font=\footnotesize, legend cell align=left},
		enlargelimits=false,
}}

\usetikzlibrary{backgrounds}

\usepackage{babel}
\RequirePackage[blocks]{authblk}
\makeatletter
\let\ps@plain\ps@empty
\def\@xivpt{14pt}
\def\@sect#1#2#3#4#5#6[#7]#8{%
  \ifnum #2<2
    \null\par\vskip-15pt
  \fi
  \ifnum #2>\c@secnumdepth
    \let\@svsec\@empty
  \else
    \refstepcounter{#1}%
    \protected@edef\@svsec{%
      \ifnum #2<4
        \hb@xt@10mm{\csname the#1\endcsname}\relax
      \else
        \hb@xt@12mm{\csname the#1\endcsname}\relax
      \fi}%
  \fi
  \@tempskipa #5\relax
  \ifdim \@tempskipa>\z@
    \begingroup
      #6{%
        \@hangfrom{\hskip #3\relax\@svsec}%
          \interlinepenalty \@M #8\@@par}%
    \endgroup
    \csname #1mark\endcsname{#7}%
    \addcontentsline{toc}{#1}{%
      \ifnum #2>\c@secnumdepth \else
        \protect\numberline{\csname the#1\endcsname}%
      \fi
      #7}%
  \else
    \def\@svsechd{%
      #6{\hskip #3\relax
      \@svsec #8}%
      \csname #1mark\endcsname{#7}%
      \addcontentsline{toc}{#1}{%
        \ifnum #2>\c@secnumdepth \else
          \protect\numberline{\csname the#1\endcsname}%
        \fi
        #7}}%
  \fi
  \@xsect{#5}}
\renewcommand\LARGE{\@setfontsize\LARGE{16}{20}}
\def\abstract#1{\def\@abstract{#1}}
\def\abstractEn#1{\def\@abstractEn{#1}}
\def\titleEn#1{\def\@titleEn{#1}}
\def\@maketitle{%
  \newpage
  \null
  \let \footnote \thanks
    {\LARGE\bfseries\RaggedRight \@title \par}%
    \vskip 1\baselineskip%
    {\normalsize
      \@author\par}%
    \vskip 2\baselineskip%
    \vskip \baselineskip%
    {\section*{Abstract}
      \@abstract}%
  \par
  \vskip 3\baselineskip}

\renewcommand\section{\@startsection {section}{1}{\z@}%
                                   {-3.5ex \@plus -1ex \@minus -.2ex}%
                                   {\baselineskip}%
                                   {\normalfont\Large\bfseries\RaggedRight}}
\renewcommand\subsection{\@startsection{subsection}{2}{\z@}%
                                     {\baselineskip}%
                                     {1ex}%
                                     {\normalfont\large\bfseries\RaggedRight}}
\renewcommand\subsubsection{\@startsection{subsubsection}{3}{\z@}%
                                     {1\baselineskip}%
                                     {3bp}%
                                     {\normalfont\normalsize\bfseries\RaggedRight}}
\renewcommand\paragraph{\@startsection{paragraph}{4}{\z@}%
                                    {1\baselineskip\@plus1ex \@minus.2ex}%
                                    {3bp}%
                                    {\normalfont\normalsize\RaggedRight}}
\renewcommand\subparagraph{\@startsection{subparagraph}{5}{\parindent}%
                                       {3.25ex \@plus1ex \@minus .2ex}%
                                       {-1em}%
                                      {\normalfont\normalsize\bfseries\RaggedRight}}
\parindent\p@
\parskip\z@
\DeclareCaptionLabelSeparator{enskip}{\enskip}
\makeatother

\newcommand{\La}{L_\mathrm{sa}}
\newcommand{\chirpbandwidth}{B}
\newcommand{\chirpslope}{\gamma}
\newcommand{\wavelength}{\lambda_\mathrm{c}}
\newcommand{\Rzd}{r_{0}}

\newcommand{\cres}{\delta_\mathrm{c}}
\newcommand{\rres}{\delta_\mathrm{r}}
\newcommand{\creshalf}{\delta_\mathrm{c,\SI{3}{\decibel}}}
\newcommand{\lightspeed}{c}

\newcommand{\winhpbwr}{\eta_\mathrm{r}}

\newcommand{\Nt}{N_\mathrm{T}}
\newcommand{\Nr}{N_\mathrm{R}}
\newcommand{\Nf}{N_{\mathrm{f}}}

\newcommand{\Nc}{N_\mathrm{c}}
\newcommand{\nslow}{n_\mathrm{s}}
\newcommand{\nfast}{n_\mathrm{f}}

\newcommand{\fstart}{f_{0}}
\newcommand{\Tchirp}{T_\mathrm{c}}
\newcommand{\Tr}{T_\mathrm{r}}

\newcommand{\Tff}{T_\mathrm{f}}
\newcommand{\fs}{f_\mathrm{s}}

\newcommand{\vego}{v_\mathrm{ego}}
\newcommand{\Gint}{G_\mathrm{int}}
\newcommand{\intexp}{\alpha}

\newcommand{\sbb}[2]{s_{#1}\left(#2\right)}
\newcommand{\SRC}[2]{S_{#1}\left(#2\right)}
\newcommand{\dftn}[1]{\operatorname{DFT}{\left\{#1\right\}}}

\newcommand{\window}[2]{w_\mathrm{#1}\left(#2\right)}
\newcommand{\Window}[2]{W_\mathrm{#1}\left(#2\right)}

\newcommand{\twr}[2]{r_{#1}\left(#2\right)}

\newcommand{\norm}[1]{\lVert#1\rVert_{2}}

\renewcommand{\vec}[1]{\mathbf{#1}}
\newcommand{\uvec}[1]{{_\mathrm{n}}\vec{u}_{#1}}
\newcommand{\virtpos}[1]{{_\mathrm{n}}\vec{p}_{#1}}
\newcommand{\targetpos}{{_\mathrm{n}}\vec{p}_{\mathrm{t}}}
\newcommand{\virtposarg}[2]{\virtpos{#1}\left(#2\right)}
\newcommand{\reals}{\mathbb{R}}

\newcommand{\grid}{\mathcal{G}}
\newcommand{\gridpoint}{{_\mathrm{n}}\vec{g}}
\newcommand{\twrgrid}[2]{\tilde{r}_{#1}\left(#2\right)}

\newcommand{\matchedfilter}[1]{H\left(#1\right)}

\newcommand{\fvib}{f_\mathrm{vib}}
\newcommand{\Avib}{A_\mathrm{vib}}
\newcommand{\xvib}{x_\mathrm{vib}}
\newcommand{\Lvib}{L_\mathrm{vib}}

\title{Indoor Synthetic Aperture Radar Measurements of Point-Like Targets Using a Wheeled Mobile Robot}

\author[a]{Yuma E. Ritterbusch}
\author[a]{Johannes Fink}
\author[b]{Christian Waldschmidt}
\affil[a]{Robert Bosch GmbH, Corporate Sector Research and Advance Engineering, 71272 Renningen, Germany}
\affil[b]{Institute of Microwave Engineering, Ulm University, 89081 Ulm, Germany}

\abstract{
Small, low-cost radar sensors offer a lighting independent sensing
capability for indoor mobile robots that is useful for localization and mapping.
Synthetic aperture radar (SAR) offers an attractive way to
increase the angular resolution of small radar sensors for use on mobile robots to generate
high-resolution maps of the indoor environment.
This work quantifies the maximum synthesizable aperture length of our
mobile robot measurement setup using radar-inertial odometry localization
and offers insights into challenges for robotic millimeter-wave SAR imaging.
}

\begin{document}
\bibliographystyle{ieeetr}
\maketitle

\section{Introduction}
\label{sec:introduction}

\begin{figure}
	\centering\includegraphics[width=0.8\columnwidth]{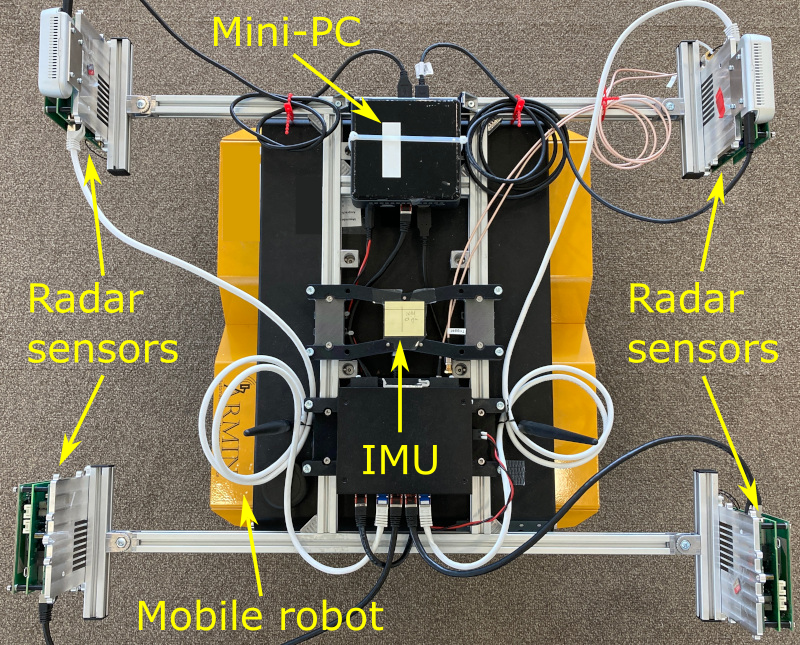}
	\caption{Remote-controlled mobile robot with radar sensors, IMU and mini-PC for data recording.}
	\label{fig:jackal}
\end{figure}

Low-cost automotive radar sensors have been successfully applied in mobile robotics
research to aid in indoor and outdoor navigation using radar-inertial odometry (RIO) \cite{doer_ekf_2020}.
These sensors have also been used for simultaneous localization and mapping (SLAM)
based on radar point clouds created using digital beamforming (DBF) \cite{almalioglu_milli-rio:_2021}.
The achievable angular resolution of DBF is directly
related to the physical extent of the multiple-input multiple-output (MIMO) antenna aperture of the sensor.
The physical antenna aperture size can only be increased by designing a larger sensor with many channels,
thus increasing manufacturing costs and power consumption.
SAR imaging techniques offer an attractive digital signal processing alternative to
achieve high angular resolution using a small antenna aperture with only a few channels.
SAR processing has previously been applied in the automotive context to image the driving environment
\cite{gao_mimo-sar:_2021}.
Another key benefit of the coherent processing of multiple radar measurements
is an improvement of signal-to-noise ratio (SNR) in the image which can help
to identify objects with smaller radar cross-section (RCS) and improve the accuracy
of high-resolution probabilistic maps in an automotive setting \cite{grebner_probabilistic_2023}.
Our focus is to bring the benefits of SAR imaging to the field
of mobile robotics in indoor scenarios.
Indoor mobile robots need to perform localization without relying
on highly accurate external
reference systems such as a global navigation satellite system (GNSS),
which limits the accuracy of the estimated robot trajectory.
Furthermore radar-inertial sensor fusion approaches \cite{doer_ekf_2020}
rely on odometry measurements only and hence suffer from the trajectory estimate drifting away from the ground truth over time.
This imperfect trajectory estimate presents a key limiting factor
to the maximum achievable synthetic aperture length.
This paper aims to quantify this limitation in our experimental setup
shown in \textbf{\cref{fig:jackal}}.
To this end we analyze the achieved cross-range resolution of SAR images
computed based on a radar-inertial odometry (RIO) trajectory estimate \cite{doer_ekf_2020}.
This contrasts with the contributions \cite{gao_mimo-sar:_2021,grebner_probabilistic_2023,harrer_synthetic_2017}
which do not explicitly mention achieved SAR resolution for non-simulated datasets.
The remainder of this paper is organized as follows. The system model and experimental setup are described in \cref{sec:system_model,sec:experimental_setup} respectively.
Results are presented and discussed in \cref{sec:results} while \cref{sec:summary} provides a summary of the findings.

\section{System model}
\label{sec:system_model}

\subsection{Signal model}

\begin{figure}
	\centering
	\includegraphics{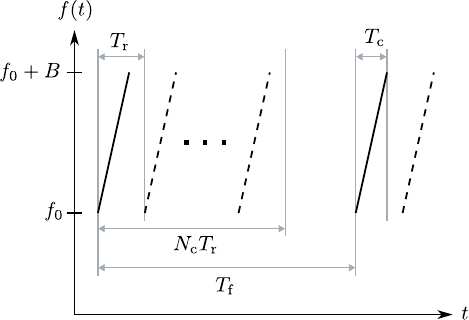}
	\caption{Chirp-sequence FMCW TDM timing diagram for $\Nt=2$ transmitters (solid and dashed lines).}
	\label{fig:timing_diagram}
\end{figure}

The radar sensors considered in this work use a chirp-sequence (CS) FMCW time-division multiplexing (TDM) MIMO modulation scheme,
in which radar frames consisting of a sequence of $\Nc$ linear frequency chirps with duration $\Tchirp$ and bandwidth $\chirpbandwidth$ are transmitted at a period $\Tff$.
During transmission of a frame, the time division multiplexing scheme cycles through $\Nt$ transmit channels at an interval $\Tr$ creating the interleaved transmit pattern shown in \textbf{\cref{fig:timing_diagram}}. The $\Nr$ receive channels are always active. Overall this results in $\Nt\Nr$ virtual channels with $\Nc/\Nt$ slow-time and $\Tchirp\fs$ fast-time samples available for baseband signal processing \cite{li_signal_2021}.
The equivalent complex baseband signal of a single point-target resulting from one chirp-sequence after downconversion and low-pass filtering is
\begin{align}
&\forall i \in \left[0,1,\dots,\Nt-1\right], j \in  \left[0,1,\dots,\Nr-1\right], \nonumber \\
&  \nslow \in \left[0,1,\dots,\Nc/\Nt-1\right], \nfast \in \left[0,1,\dots,\Tchirp\fs -1\right]: \nonumber \\
& \sbb{ij}{\nslow,\nfast} = A_{ij}  \exp \left\{ j \frac{2\pi}{\lightspeed} \left( \fstart + \chirpslope \frac{\nfast}{\fs} \right) \twr{ij}{\nslow,\nfast}  \right\}
\label{eq:basebandsignal}
\end{align}
with amplitude $A_{ij}$, chirp-rate $\chirpslope = \chirpbandwidth / \Tchirp$,
and the propagation speed of electromagnetic waves $\lightspeed$ in the considered medium.
The two-way target range for the pair of transmitter $i$ and receiver $j$ is
\begin{align}
	\twr{ij}{\nslow,\nfast} =& \norm{\virtposarg{i}{\nslow,\nfast} - \targetpos} \nonumber \\ & \quad + \norm{\virtposarg{j}{\nslow,\nfast} - \targetpos},
\label{eq:twowaytargetrange}
\end{align}
where the vectors $\virtpos{i}, \virtpos{j}, \targetpos \in \reals^3$ denote
the position vectors of transmit channel $i$, receive channel $j$ and the target
respectively, expressed in common reference frame $\mathrm{n}$. $\norm{\cdot}$ denotes the Euclidean distance.

\subsection{SAR image formation}

\begin{figure}
	\centering
	\includegraphics{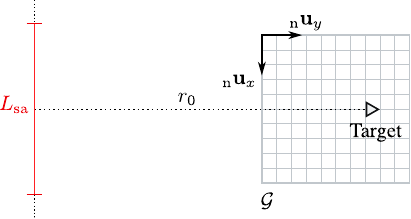}
	\caption{Side-looking imaging geometry showing the synthetic aperture highlighted in red. The image grid surrounding the target of interest is defined
		by the basis vectors  $\uvec{x}$ and $\uvec{y}$.}
	\label{fig:synthetic_aperture}
\end{figure}

The SAR images are computed using a time-domain backprojection (TDBP) algorithm \cite{harrer_synthetic_2017},
which first performs a range-compression of the baseband signal by applying a window function $\window{r}{\nfast}$ for sidelobe control and DFT along the fast-time index $\nfast$ to obtain the range profiles
\begin{align}
	&\SRC{ij}{\nslow,r} = \dftn{\window{r}{\nfast}\sbb{ij}{\nslow,\nfast}} \nonumber \\
		& \, \approx  A_{ij} \exp \left\{ j \frac{2\pi\fstart}{\lightspeed} \twr{ij}{\nslow} \right\} \nonumber \\ &\qquad\qquad
		\cdot \Window{r}{\frac{2\pi\chirpslope}{\lightspeed\fs}\left(r - \twr{ij}{\nslow}\right)},
\label{eq:rangeprofile}
\end{align}
where the range-change during the short chirp duration $\Tchirp$ is assumed to be sufficiently small
such that the fast-time index $\nfast$ in the exponential can be omitted. $\Window{r}{\Omega}$ is the DTFT of the window function $\window{r}{\nfast}$.

The range-resolution after range-compression is \cite{richards_principles_2010}
\begin{align}
\rres = \winhpbwr \frac{\lightspeed}{2 \chirpbandwidth},
\label{eq:range_resolution}
\end{align}
with the window-dependent main lobe-broadening factor $\winhpbwr$.
The second step of SAR image formation is the azimuth-compression.
The complex amplitudes of each range profile $\SRC{ij}{\nslow,r}$
are interpolated onto an image grid
$\grid \subset \reals^3$ and a matched filter
\begin{align}
	\matchedfilter{r} = \exp \left\{-j \frac{2\pi\fstart}{\lightspeed} r \right\}
\end{align}
is applied for every grid point.
Finally all available range-profile samples are coherently integrated to yield the image
\begin{align}
	I\left(\grid\right) = \sum_{i=0}^{\Nt-1} \sum_{j=0}^{\Nr-1} \sum_{\nslow=0}^{\frac{\Nc}{\Nt}-1}
	& \SRC{ij}{\nslow, \twrgrid{ij}{\nslow}} \matchedfilter{\twrgrid{ij}{\nslow}},
\label{eq:sarimage}
\end{align}
where
\begin{align}
\forall \gridpoint \in \grid: \quad
	\twrgrid{ij}{\nslow} =\, & \norm{\virtposarg{i}{\nslow}-\gridpoint} \nonumber \\ & \quad + \norm{\virtposarg{j}{\nslow}-\gridpoint}
\label{eq:twowaygridrange}
\end{align}
is the two-way range from each transmitter/receiver pair $ij$ to every point
in the image grid.
\textbf{\Cref{fig:synthetic_aperture}} shows the SAR imaging geometry
for a synthetic aperture of length $\La$ and a range of closest
approach to the target $\Rzd$.
The cross-range resolution $\cres$ of the resulting SAR image after azimuth-compression is \cite{moreira_tutorial_2013}
\begin{align}
\cres = \frac{\wavelength}{2 \La} \Rzd,
\label{eq:cross_range_resolution}
\end{align}
where $\wavelength$ is the wavelength at the chirp center frequency.
The expression \cref{eq:cross_range_resolution} gives the distance between
the main lobe peak and first null, whose position is difficult to determine
in noisy measurements.
Instead we use the half-power main lobe width
\begin{align}
	\creshalf = 0.88448 \, \cres
	\label{eq:cross_range_halfpower}
\end{align}
to determine the achieved resolution,
where the constant factor is due to the implicit rectangular window used during azimuth compression \cite{doerry_catalog_2017}.

The integration gain $\Gint$ describes the
SNR increase in the SAR image achieved by coherent processing.
When integrating $\Nf$ samples, the integration gain can be expressed as \cite{richards_principles_2010}
\begin{align}
	\Gint = \Nf^\intexp.
	\label{eq:integration_gain}
\end{align}
The exponent $\intexp$ is a measure of the integration efficiency.
In the case of coherent integration $\intexp = 1$.
In case of non-coherent integration the integration efficiency is reduced
such that $0.5 \leq \alpha < 1$ \cite{richards_principles_2010}.

\section{Experimental setup}
\label{sec:experimental_setup}
\begin{table}
	\centering
	\begin{tabular}{l c c}
		\toprule
		\textbf{Parameter}	&	\textbf{Symbol}	& \textbf{Value} \\
		\midrule
		Start frequency	&	$\fstart$	& \SI{76}{\giga\hertz} \\
		Modulation bandwidth	&	$\chirpbandwidth$ & \SI{4}{\giga\hertz} \\
		Chirp duration & $\Tchirp$ & \SI{180}{\micro\second} \\
		Chirp repetition interval & $\Tr$ & \SI{200}{\micro\second} \\
		Chirps per frame & $\Nc$ & 256 \\
		Frame repetition interval & $\Tff$ & \SI{52}{\milli\second} \\
		Baseband sample rate & $\fs$ & \SI{977}{\kilo\hertz} \\
		Number of transmit channels & $\Nt$ & 4 \\
		Number of receive channels & $\Nr$ & 4 \\
		\bottomrule
	\end{tabular}
	\caption{Radar waveform parameters.}
	\label{tab:radar_parameters}
\end{table}
The experimental setup consists of four chirp-sequence FMCW MIMO radar sensors
operating between \SIlist{76;81}{\giga\hertz} which is the frequency range defined by ETSI for automotive radar applications \cite{noauthor_etsi_2017-1,noauthor_etsi_2017}.
\textbf{\Cref{tab:radar_parameters}} summarizes the important waveform parameters.
The radar sensors are mounted onto a remote controlled mobile robot facing outward
and perpendicular to the direction of movement. This enables SAR imaging of the surrounding scene.
In addition to the radar sensors, an inertial measurement unit (IMU) is also attached to the center of rotation of the robot.
The start of the IMU sample recording is synchronized to the first chirp sequence of a data collection run by a hardware trigger signal.
The chirp-sequences of all radar sensors are synchronized via Ethernet using the precision time protocol (PTP). To avoid interference between
the simultaneously operating sensors, their start frequencies were
offset relative to each other by \SI{100}{\mega\hertz}. \Cref{fig:jackal} shows the mobile robot with the mounted radar sensors used to record the dataset.

\begin{figure}
	\centering
	\includegraphics[width=0.6\columnwidth]{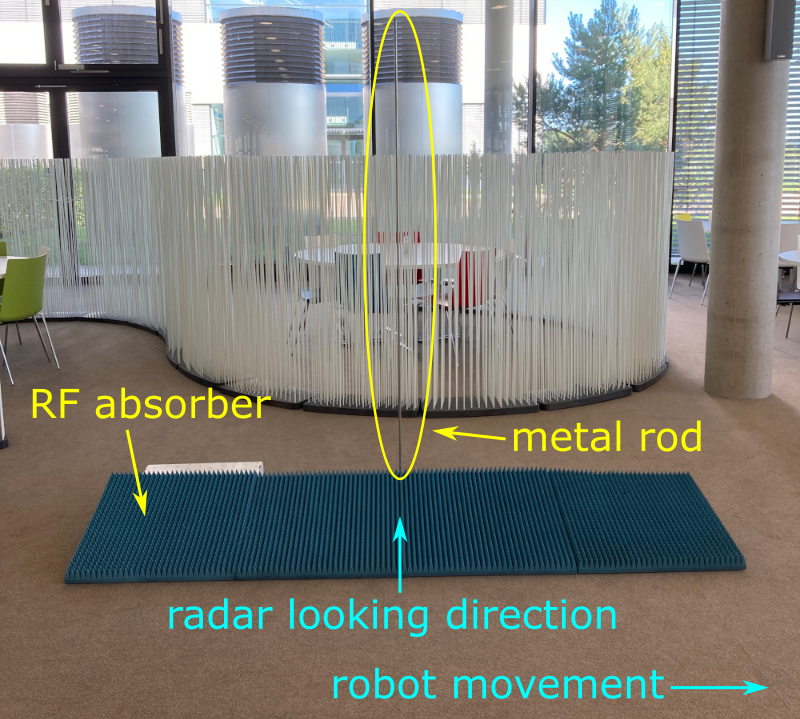}
	\caption{Vertical metal rod (inside ellipse), RF absorber and side-looking SAR imaging geometry.}
	\label{fig:metal_rod}
\end{figure}

Two point-like reference targets were placed in the scene to determine the achieved SAR image resolution based on their main lobe width.
The first reference target is a metal rod with a diameter of \SI{10}{\milli\meter} and a length of \SI{2}{\meter}, held vertically by a wooden block which
was covered in RF absorbers to suppress strong unwanted reflections as shown in \textbf{\cref{fig:metal_rod}}.
The second reference target is a trihedral corner reflector (TCR) which was directly placed on top of pyramidal RF absorbers.
During data collection the robot was remotely controlled to maintain a straight line path at constant speed $\vego$ while passing the reference targets.
This results in the side-looking imaging geometry shown in \cref{fig:synthetic_aperture}.
After data collection the robot trajectory is estimated using radar-inertial odometry (RIO) as described in~\cite{doer_ekf_2020}.
The transmit and receive channel positions $\virtpos{i}$ and $\virtpos{j}$ are computed from the estimated robot trajectory using the known mounting position, orientation, and antenna array geometry of each radar for every sample of the
coherent processing interval. For simplicity, the SAR images used for the performance evaluation were computed from
a single radar sensor only, while data from all four radars was used for trajectory estimation.

\section{Results}
\label{sec:results}

\subsection{Range resolution}

\begin{figure}
\includegraphics{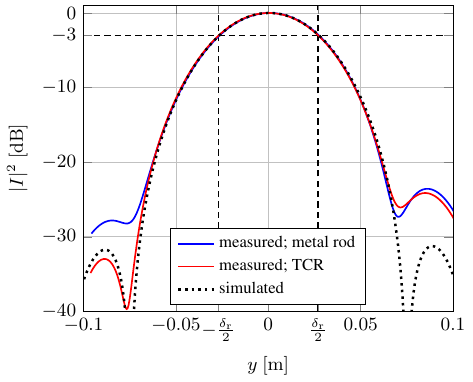}
\caption{Measured range peak profiles of the metal rod, trihedral corner reflector (TCR) and simulated point-target for a Hann window ($\winhpbwr = 1.4381$ \cite{doerry_catalog_2017}).}
\label{fig:range_profile}
\end{figure}

\textbf{\Cref{fig:range_profile}} shows the range profile of the metal rod, trihedral corner reflector and a point-target simulation using the radar waveform parameters from \cref{tab:radar_parameters}.
The vertical dashed lines mark the half-power main lobe width as predicted
by \cref{eq:range_resolution} for the Hann window used during range compression  ($\winhpbwr=1.4381$ \cite{doerry_catalog_2017}), indicating that the computed SAR image has achieved the range resolution to be expected from the used modulation bandwidth.

\subsection{Cross-range resolution}

In order to determine the achievable cross-range resolution, the same
dataset was used to compute SAR images  with successively greater synthetic
aperture lengths $\La$.
\textbf{\Cref{fig:cross_range_resolution_rod,fig:cross_range_resolution_tcr}} show the resulting achieved cross-range resolution
for the reference targets metal rod and trihedral corner reflector.
Each diagram depicts data from two different measurement passes performed at robot speeds $\vego$ of \SI{0.4}{\meter\per\second} and \SI{1.0}{\meter\per\second}. The dashed line indicates the cross-range half-power main lobe width $\creshalf$ as predicted by \cref{eq:cross_range_halfpower} for both scenarios.

\subsubsection{Metal rod}
\begin{figure}[t]
	\centering
	\includegraphics{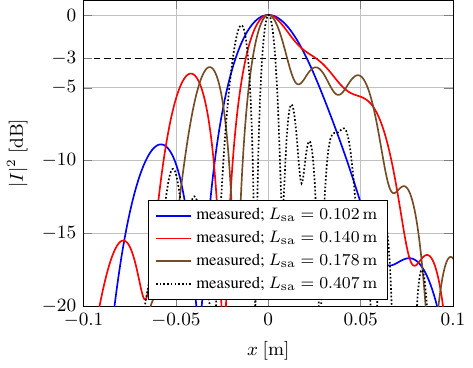}
	\caption{Measured metal rod cross-range peak profiles for $\vego=\SI{0.4}{\meter\per\second}$.}
	\label{fig:cross_range_profile_rod}
\end{figure}

In case of the metal rod a good agreement between expected and achieved main lobe width
is apparent for synthetic aperture lengths $\La \leq \SI{0.34}{\meter}$.
Because the range of closest approach $\Rzd$ for the depicted passes is \SI{2}{\meter},
synthetic apertures longer than $\SI{0.34}{\meter}$ lead to an expected cross-range resolution that is smaller than the diameter of the metal rod.
As the apertures become larger a splitting of the single target peak into two peaks of
almost equal power can be observed.
An example of this phenomenon is shown with the black dotted line in \textbf{\cref{fig:cross_range_profile_rod}}.
This is likely caused by the scattering
center on the surface of the metal rod shifting location as the radar moves past.
It seems unlikely that this is caused by a creeping wave around the surface of the metal rod,
as its received power is much smaller than that of the primary reflection.
Furthermore, the additional propagation delay due to the small rod diameter is not resolvable
at the used modulation bandwidth.

When the cross-range resolution approaches the physical target size,
the point-target assumption is violated and the half-power main lobe
criterion no longer seems suitable for determining the achieved cross-range resolution.

\begin{figure}[t]
	\centering
	\includegraphics{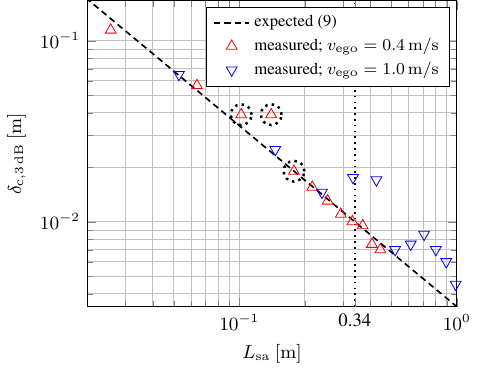}
	\caption{Achieved cross-range half-power main lobe width for metal rod.}
	\label{fig:cross_range_resolution_rod}
\end{figure}

The red triangles in \cref{fig:cross_range_resolution_rod} show the achieved resolution
for a robot speed of \SI{0.4}{\meter\per\second}. One thing to note there is an outlier around $\La = \SI{0.14}{\meter}$, while the samples immediately preceding and following agree well with the theory.
\Cref{fig:cross_range_profile_rod} shows the cross-range profiles for the three circled neighboring samples in \cref{fig:cross_range_resolution_rod}.
In case of the shortest aperture $\La=\SI{0.102}{\meter}$,
the cross-range spectrum exhibits a well defined peak. The outlier sample
on the other hand has a much broader peak and a shoulder that hints at a side lobe,
causing the inflated estimate for achieved cross-range resolution.
The third sample with $\La=\SI{0.178}{\meter}$
again exhibits a well defined peak, while a side lobe only about \SI{3.5}{\decibel} below the main peak has become visible. This new side lobe is located around the same
cross-range position as the shoulder in the previous profile.
Because there are no targets located close to the metal rod in the experimental
setup, we can conclude that this is in fact a side lobe and not an actual target reflection.
Possible sources for these side lobes with significantly higher level
than the \SI{-13.3}{\decibel} to be expected from a rectangular window \cite{doerry_catalog_2017} will be discussed in \cref{sec:vibration}.

The blue triangles in \cref{fig:cross_range_profile_rod} show the achieved resolution for a robot speed of \SI{1}{\meter\per\second}. Similar to the lower speed case the results follow the
expected resolution until a synthetic aperture length of $\La\leq\SI{0.34}{\meter}$.
For longer synthetic apertures the deviation from the theory becomes quite large,
again indicating problems related to a violation of the single point-target assumption.


\subsubsection{Trihedral corner reflector}

\begin{figure}[t]
	\includegraphics{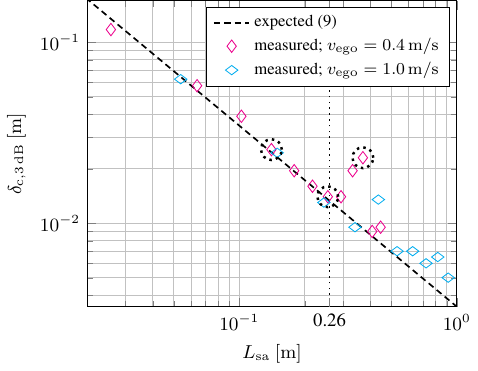}
	\caption{Achieved cross-range half-power main lobe width for trihedral corner reflector.}
	\label{fig:cross_range_resolution_tcr}
\end{figure}

\begin{figure}[t]
	\includegraphics{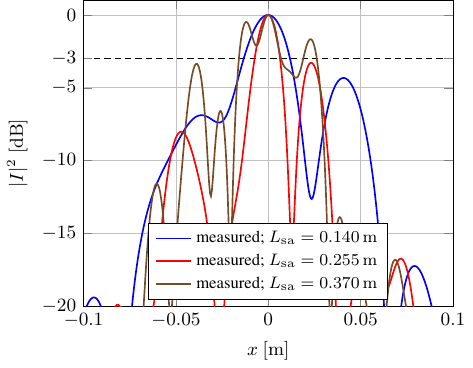}
	\caption{Measured trihedral corner reflector cross-range peak profiles for $\vego=\SI{0.4}{\meter\per\second}$.}
	\label{fig:cross_range_profile_tcr}
\end{figure}

The results for the trihedral corner reflector shown in \cref{fig:cross_range_resolution_tcr} are very similar to those
of the metal rod previously discussed.
For a robot speed of \SI{0.4}{\meter\per\second} depicted by the magenta diamonds,
a good agreement with the expected cross-range half-power main lobe width can be observed
until a synthetic aperture length of $\La = \SI{0.26}{\meter}$.
For larger aperture lengths a strong deviation from the expected
main lobe width is visible.
\textbf{\Cref{fig:cross_range_profile_tcr}} shows the cross-range profile for the highlighted aperture lengths in
\cref{fig:cross_range_resolution_tcr}. In case of $\La=\SI{0.14}{\meter}$ there
is a well-defined main peak, albeit with an already high side lobe level of $\SI{-4.5}{\decibel}$. As the aperture length increases the main lobe width
decreases while side lobe level increases. In case of $\La=\SI{0.37}{\meter}$,
the main lobe appears to have split apart and the half-power main lobe width
is much wider than expected.

For a robot speed of \SI{1}{\meter\per\second}, a slightly longer aperture of $\La=\SI{0.34}{\meter}$
can be synthesized before the achieved cross-range resolution no longer tracks the expected behavior.
Looking at the side lobe structure in \cref{fig:cross_range_profile_tcr} it seems like the side lobes
are the major limiting factor in our SAR imaging setup.


\subsection{Sensor vibration}
\label{sec:vibration}

\begin{figure}
\includegraphics{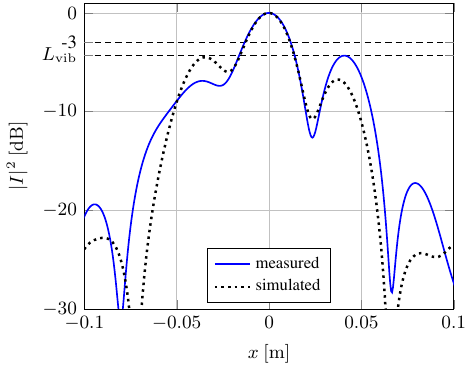}
\caption{Measured trihedral corner reflector and simulated point-target cross-range peak profiles for a sensor vibration with $\fvib = \SI{3.83}{\hertz}$ and $\Avib=\SI{0.319}{\milli\meter}$.}
\label{fig:cross_range_profile_vibration}
\end{figure}

The high side lobe level can be plausibly explained by low-frequency
vibration of the radar sensor during data recording, which is known
to cause side lobes in the Doppler-spectrum.
The Doppler-spectrum resulting from a sinusoidal motion of the radar sensor
during the coherent processing interval can be written
in terms of an infinite series of Bessel-functions \cite{hau_degradation_2020}.
\textbf{\Cref{fig:cross_range_profile_vibration}} shows the measured trihedral corner reflector cross-range profile for $\La=\SI{0.14}{\meter}$
and the results of a point-target simulation with a sinusoidal vibration of the radar sensor along the $y$-direction (sensor boresight) during data recording.
The simulated robot speed was $\vego=\SI{0.368}{\meter\per\second}$ to achieve the same aperture length $\La$ as in the measurement.
In order to parametrize the simulation, a single dominant
vibration component was assumed to be present. The vibration frequency $\fvib$ was chosen based
on the expected Doppler-frequency of a target at side lobe position $\xvib$
\begin{align}
	\fvib = \frac{2 \vego}{\wavelength} \frac{\xvib}{\sqrt{\xvib^2 + \Rzd^2}},
\end{align}
which for $\xvib=\SI{0.04}{\meter}$, $\Rzd=\SI{2}{\meter}$ yields $\fvib = \SI{3.83}{\hertz}$.
The vibration amplitude $\Avib$ was chosen based on the side lobe-level $\Lvib$ such
that it fulfills
\begin{align}
	\Lvib = 20 \log_{10} \frac{J_1\left(a\right)}{J_0\left(a\right)},
\end{align}
where
\begin{align}
	a = 2\pi \frac{2\Avib}{\wavelength}
\end{align}
is derived in \cite{hau_degradation_2020} and $J_i$ is the $i$-th order Bessel-function
of the first kind.
For $\Lvib = \SI{-4.3}{\decibel}$ this yields $\Avib=\SI{0.319}{\milli\meter}$.
As to be expected, the highly simplified simulated vibration does not
perfectly match the measurement result, it does however cause
the same characteristic paired and slightly
asymmetric side lobes to appear next to the main lobe.
Based on this simulation it seems plausible that vibration of the radar sensor
in our measurement setup causes the image degradation in terms of side lobe level.
The primary focus of this paper is to quantify the performance limitations of our specific mobile robot-based SAR system.
Therefore low-vibration measurement setups, such as step-motor driven linear rail, were not investigated in this study.
Furthermore it was not investigated whether the vibration-induced sidelobe level
could be reduced using autofocus algorithms.

\subsection{Integration gain}

\begin{figure}
	\centering
	\includegraphics{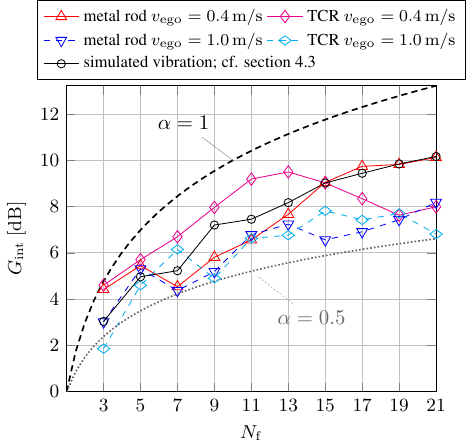}
	\caption{Measured integration gain for different number of integrated frames. The dashed line shows the upper limit, while the gray dotted line shows the lower limit for non-coherent integration according to \cref{eq:integration_gain}.}
	\label{fig:integration_gain}
\end{figure}

\textbf{\Cref{fig:integration_gain}} shows the achieved integration gain for all previously presented measurement passes
as a function of the number of integrated radar frames $\Nf$.
Unsurprisingly, the real-world scenarios fall short of the theoretical maximum
coherent integration gain predicted by \cref{eq:integration_gain}, shown as a dashed line.
For comparison, the integration gain of a simulated point-target scenario with a sinusoidal vibration as in \cref{sec:vibration}
is shown in black. The simulation results were averaged over 7 individual simulations with different starting phases.
No averaging was applied to the measured results. The simulation results further support the hypothesis that
even with otherwise perfect knowledge of the trajectory,
vibration of the radar sensor negatively affects the integration gain.
In case of the metal rod depicted by the triangle markers, the integration gain follows the general shape of
the simulation with a similar dip at $\Nf=7$.
The performance in terms of achieved gain is also worse than the simulation.
This seems plausible, as the real measurement scenario likely includes more complex vibration patterns
and unlike the simulation also suffers from errors in trajectory estimates.
For the case of the trihedral corner reflector denoted by the diamonds, the integration gains drop for
larger synthetic apertures, further indicating problems related to drifting trajectory estimates.
Unlike non-coherent integration, the SAR image formation in \cref{eq:sarimage}
always operates on complex valued data. In the worst case this can lead to destructive interference of correlated target returns
when trajectory estimates include sufficiently large errors. As the noise is assumed to be uncorrelated on the other hand,
it should not be affected by trajectory drift during integration. The overall result is a decrease in integration gain.
At a higher robot speed of $\vego=\SI{1}{\meter\per\second}$, depicted by the dashed lines,
the same number of integrated frames corresponds to a longer synthetic aperture.
This means that the accumulated trajectory errors are also larger, explaining
worse performance than in the lower speed case.

\section{Summary}
\label{sec:summary}

We have shown that SAR imaging using automotive radar chirp-sequence FMCW MIMO radar sensors mounted to a mobile robot based on radar-inertial odometry trajectory estimates is possible. The imaging results agree with expectations from
well-known theoretical results for non-degenerate imaging cases.
Vibration of the radar sensor during data collection has been identified as
a major factor degrading imaging performance in terms of
achievable side lobe level and integration gain.
Future work will focus on improving image quality to
aid in SAR-based mapping of the indoor environment.
One obvious improvement to reduce sensor vibration is
to increase the mechanical stiffness of the radar mounts and
avoid suspending the sensor weight from a long lever arm.

\bibliographystyle{abbrv}
\bibliography{references}
\end{document}